\documentclass[aps,letterpaper,twocolumn,prl,footinbib,floatfix]{revtex4}
\usepackage{amsmath}
\usepackage{amsfonts}
\usepackage{amssymb}
\usepackage[scanall]{psfrag}
\usepackage{psfrag}
\usepackage{graphicx}
\usepackage{color}

\begin{document}
\newcommand{\of}[1]{\left( #1 \right)}
\newcommand{\sqof}[1]{\left[ #1 \right]}
\newcommand{\abs}[1]{\left| #1 \right|}
\newcommand{\avg}[1]{\left< #1 \right>}
\newcommand{\cuof}[1]{\left \{ #1 \right \} }
\newcommand{\bra}[1]{\left < #1 \right | }
\newcommand{\ket}[1]{\left | #1 \right > }
\newcommand{\pil}{\frac{\pi}{L}}
\newcommand{\bx}{\mathbf{x}}
\newcommand{\by}{\mathbf{y}}
\newcommand{\bk}{\mathbf{k}}
\newcommand{\bp}{\mathbf{p}}
\newcommand{\bl}{\mathbf{l}}
\newcommand{\bq}{\mathbf{q}}
\newcommand{\bs}{\mathbf{s}}
\newcommand{\psibar}{\overline{\psi}}
\newcommand{\svec}{\overrightarrow{\sigma}}
\newcommand{\dvec}{\overrightarrow{\partial}}
\newcommand{\bA}{\mathbf{A}}
\newcommand{\bdelta}{\mathbf{\delta}}
\newcommand{\bK}{\mathbf{K}}
\newcommand{\bQ}{\mathbf{Q}}
\newcommand{\bG}{\mathbf{G}}
\newcommand{\bw}{\mathbf{w}}
\newcommand{\bL}{\mathbf{L}}
\newcommand{\ohat}{\widehat{O}}
\newcommand{\up}{\uparrow}
\newcommand{\down}{\downarrow}
\newcommand{\MM}{\mathcal{M}}
\author{Eliot Kapit}
\affiliation{Department of Physics and Engineering Physics, Tulane University, New Orleans, LA 70118}
\affiliation{Initiative for Theoretical Science, The Graduate Center, City University of New York, New York, NY 10016}

\title{A Very Small Logical Qubit}

\begin{abstract}

Superconducting qubits are among the most promising platforms for building a quantum computer. However, individual qubit coherence times are not far past the scalability threshold for quantum error correction, meaning that millions of physical devices would be required to construct a useful quantum computer. Consequently, further increases in coherence time are very desirable. In this letter, we blueprint a simple circuit consisting of two transmon qubits and two additional lossy qubits or resonators, which is passively protected against all single qubit quantum error channels through a combination of continuous driving and engineered dissipation. Photon losses are rapidly corrected through two-photon drive fields implemented with driven SQUID couplings, and dephasing from random potential fluctuations is heavily suppressed by the drive fields used to implement the multi-qubit Hamiltonian. Comparing our theoretical model to published noise estimates from recent experiments on flux and transmon qubits, we find that logical state coherence could be improved by a factor of forty or more compared to the individual qubit $T_1$ and $T_2$ using this technique. We thus demonstrate that there is substantial headroom for improving the coherence of modern superconducting qubits with a fairly modest increase in device complexity.

\end{abstract}

\maketitle

\section{Introduction}

A universal quantum computer could provide enormous computing power \cite{nielsenchuang}, but all attempts to construct such a device have been stymied by noise arising from uncontrolled interactions between the physical qubits and their environment. These quantum errors can be mitigated by quantum error correction \cite{kitaev2003,bombin2010,fowlersurface,yaowang,terhal2014}, where a logical bit is encoded in the collective state of a much larger number of physical quantum bits, and complex parity-check operations (stabilizers) are repeatedly measured to algorithmically detect or correct errors before they can proliferate. Unfortunately, the overhead requirements for implementing a fault-tolerant quantum code are daunting \cite{fowlersurface}. To help supplement these complex process, a growing body of work \cite{poyatoscirac1996,diehlmicheli,krausbuchler,verstraete,pastawskikay2009,pastawskiclemente2011,vollbrechtmuschik2011,murchvool2012,kastoryanowolf2013,shankarhatridge2013,aronkulkarni2014,mirrahimileghtas2014,cohenmirrahimi2014,kapithafezi2014,kapitchalker2015,gourgyramasesh2015,leghtastouzard2015} has shown that carefully tuned quantum noise, in the form of engineered dissipation, can protect states against the effects of the unwanted noise. However, these approaches introduce their own drawbacks and overhead, and finding the minimal useful implementation-- the simplest device which can be built with current technology and passively correct or suppress all single qubit quantum error channels-- has remained an elusive challenge. It is the goal of this article to blueprint such a circuit using mature, widely adopted superconducting device technologies.

Loosely inspired by recent proposals for ``cat state qubits" in superconducting resonators \cite{mirrahimileghtas2014,sunpetrenko2014,leghtastouzard2015}, and directly adapting the shadow lattice passive error correction architecture previously developed by the author and colleagues \cite{kapithafezi2014,kapitchalker2015}, we propose a logical qubit which could consist of two transmon qubit devices coupled by driven SQUIDs to each other and to one additional lossy object (either a qubit or resonator) each. By exploiting the particular noise spectra of errors in superconducting qubits, this device demonstrates that passive error correction via resonant energy transfer to a lossy system can dramatically outperform active, measurement-based error correction in small systems, with photon loss error correction rates approaching 10 MHz for realistic device parameters (in contrast to the $\sim$1 MHz rates from measurement-based methods \cite{kellybarends2015}). Further, it achieves this rapid error correction using a simpler circuit of just two primary qubit devices and two resonators, in contrast to the six or more qubits (five primary qubits and at least one ancilla qubit to facilitate stabilizer measurement) required to correct a single error of any type using the Laflamme code \cite{laflammemiquel1996}, and twenty-five qubits for a distance 3 surface code. While dephasing ($z$ noise) is not corrected by this circuit, the continuous drive fields used to implement passive error correction suppress its effects, and we will show that for decoherence rates observed in modern qubit designs, the effect of $z$ noise here will generally be weaker than that of photon losses. Further, logical gates on or between these qubits are surprisingly simple and we do not expect them to take significantly longer than gates on or between ordinary transmon qubits. 

\begin{figure}
\includegraphics[width=3.0in]{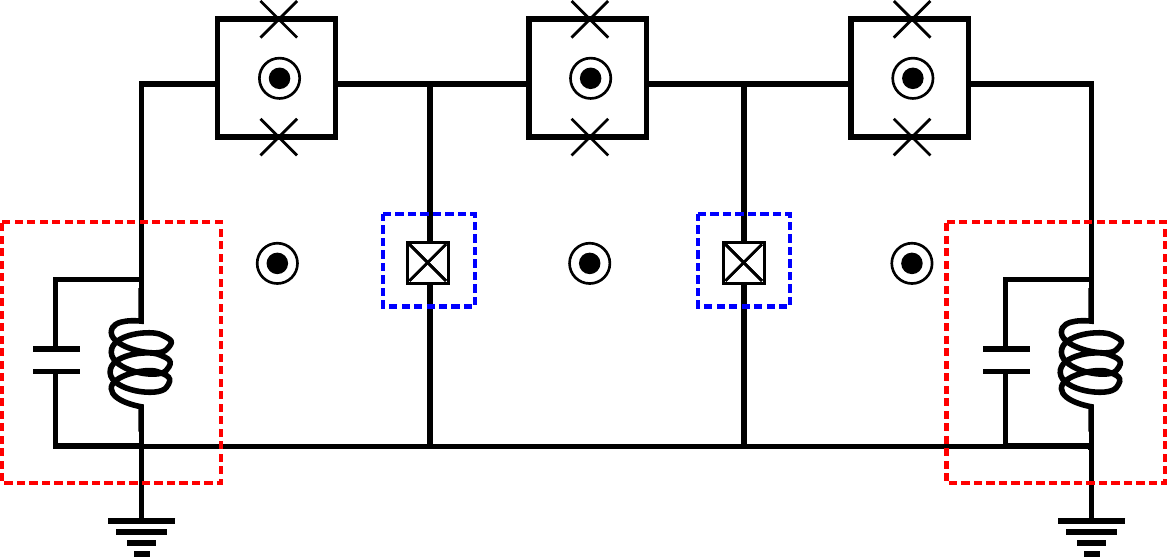}
\caption{One possible implementation of our logical qubit. The two transmon qubits (blue boxes) are the good quantum degrees of freedom we wish to protect, and the two readout resonators (red boxes) are lossy objects we will use for error correction. The three driven SQUID couplings have precisely tuned flux biases (black circles) to enable parametric interactions, as discussed in the supplemental material.}\label{circuitfig}
\end{figure}

\section{Basic Circuit Model}

For clarity and generality, we will consider a simplified theoretical model for our circuit, and leave the finer details of an example implementation and the derivation of the various terms to the supplemental material. We consider a pair of three-level superconducting qubit devices, labelled by $l$ and $r$, where the three levels correspond to device occupation by zero, one or two photons. There is a nonlinearity $-\delta$ for adding a second photon to either device compared to the $0 \to 1$ energy. We couple the two devices via a high-frequency, driven coupling which does not conserve photon number \cite{zakkabajjani2011,kapit2015,siroiscastellanos2015}, and couple each device via a similar coupling to a second, lossy degree of freedom, such as a rapidly decaying qubit or readout resonator, with a full example circuit shown in FIG.~\ref{circuitfig}. We now make the following operator definitions. We let $P_{k}^{n} \equiv \ket{n_k} \bra{n_k}$ be the projector onto all states with $n$ photons in object $k$ (and any number of photons in the other parts of the circuit). We further define $\tilde{X}_l \equiv \of{ a_l^\dagger a_l^\dagger + a_l a_l}/2$ and $\tilde{Z}_l \equiv P_l^2 - P_l^0$ (and similarly for $r$), where $a_{l}$ annihilates a photon in the left device. We now define our two-device, rotating frame Hamiltonian $H_P$ by:
\begin{eqnarray}\label{mainH}
H_P &=& - W \tilde{X}_l \tilde{X}_r  + \frac{\delta}{2} \of{P_{l}^{1} + P_{r}^{1}}.
\end{eqnarray}
$H_P$ has two ground states, $\tilde{X}_l = \tilde{X}_r = 1$ or $\tilde{X}_l = \tilde{X}_r = -1$, which we label $\ket{L_0}$ and $\ket{L_1}$ and choose to act as our logical state manifold. Note that $\tilde{X}_k \ket{1_k} = 0$ due to the three-body constraint.

We now turn to the lossy ``shadow" objects, which without loss of generality we will take to be resonators and which we label $Sl$ and $Sr$ (the $S$ label denotes a shadow object, as discussed in \cite{kapithafezi2014,kapitchalker2015}), with energies $\omega_{Sl}$ and $\omega_{Sr}$. The shadow objects are coupled to the primary qubit devices through driven couplings of a different form, yielding the final qubit-resonator Hamiltonian 
\begin{eqnarray}\label{HPS}
H_{PS} + H_{S} &=&  \of{W + \frac{\delta}{2} }  \of{ a_{Sl}^\dagger a_{Sl} +  a_{Sr}^\dagger a_{Sr} } \\ & &+ \Omega \of{a_l^\dagger a_{Sl}^\dagger + a_r^\dagger a_{Sr}^\dagger + {\rm H.c.} }. \nonumber
\end{eqnarray}
Our final device Hamiltonian is simply $H = H_P + H_{PS} + H_{S}$; the precise details of the signal configurations and wire network necessary to obtain this Hamiltonian are described in the supplemental material. We will now show that given a resonator decay rate $\Gamma_S$ which is fast compared to the photon loss rate $\Gamma_P$ $(= 1/T_1)$ in the two qubits, this circuit is passively protected against all single qubit errors, leading to exceptionally long lifetimes for the logical ground states $\ket{L_0}$ and $\ket{L_1}$.

\section{Error Correction: Photon losses}

\begin{figure}
\includegraphics[width=3.25in]{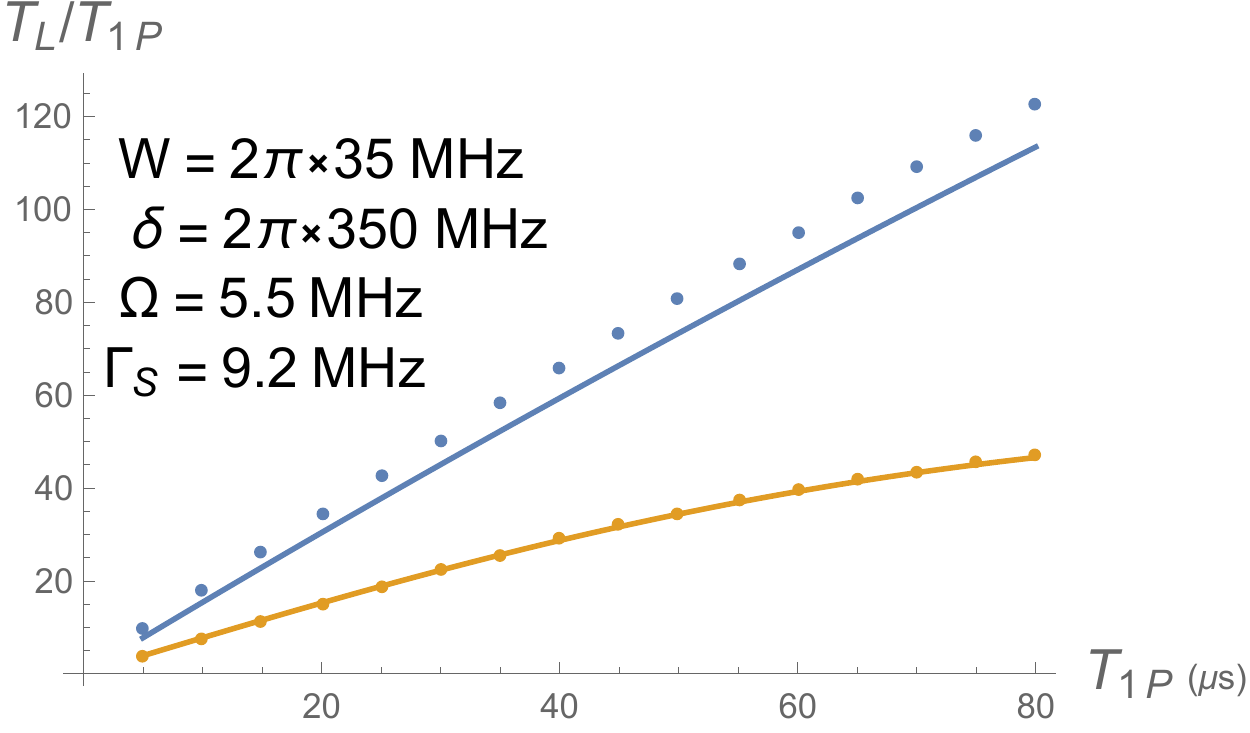}
\caption{Strong enhancement of the effective logical state lifetimes $T_{L}$ against photon losses through engineered dissipation, with all times in ${\rm \mu s}$. Each data point is computed by numerically integrating the Lindblad equations for the Hamiltonian $H = H_P + H_S + H_{PS}$ from (\ref{mainH},\ref{HPS}), with parameters as stated in the plot legend and varying ``bare" $T_{1P} \equiv 1/\Gamma_P$ from photon losses in the two primary qubits. The blue points plot the improvement factor $T_{1L}/T_{1P}$; $T_{1L}$ captures the decay of the system to an incoherent mixture of states after initialization to $\ket{0_L}$, extracted by fitting the measure ${\rm Tr }\sqof{ \rho  \ket{0_L} \bra{0_L} }  $ to an exponential decay law, with short term transient behavior dropped. The orange points plot the improvement factor $T_{2L}/T_{1P}$ of the ``dephasing" time $T_{2L}$, extracted by initializing the system to $\tilde{Z}_l \tilde{Z}_r = \pm 1$ and fitting ${\rm Tr } \sqof{\rho \tilde{Z}_l \tilde{Z}_r } $ (the plotted value is the average of the $\tilde{Z}_l \tilde{Z}_r = +1$ and $\tilde{Z}_l \tilde{Z}_r = -1$ states). The two continuous curves plot the analytically predicted improvement factor from the rates calculated in (\ref{GEminus}). The lifetime $T_{2L}$ is reduced by the constant error term in $\Gamma_{E}^X$; this term does not limit $T_{1L}$, as the system is initialized in an $\tilde{X}$ eigenstate. Increasing the nonlinearity $\delta$ would raise the limit for $T_{2L}$ improvement by passive error correction. Note that all points on the plot are past the ``breakeven" point ($T_L = T_{1P}$), which occurs around $T_{1P} \simeq 1 \mu s$.} \label{T1eff}
\end{figure}

We first tackle photon loss errors, a white noise error source which to good approximation occurs at rates independent of many-body energetics ($W$ terms). Without loss of generality, we consider a single photon loss in the left qubit, which sends:
\begin{eqnarray}
a_l \ket{L_0} \to \ket{1_l} \otimes \frac{\ket{0_r} + \ket{2_r}   }{\sqrt{2}} \otimes \ket{0_{Sl} 0_{Sr}}, \\ 
a_l \ket{L_1} \to \ket{1_l} \otimes \frac{-\ket{0_r} + \ket{2_r}   }{\sqrt{2}} \otimes \ket{0_{Sl} 0_{Sr}}. \nonumber
\end{eqnarray}
However, these states are not eigenstates of $H$, due to the qubit-resonator couplings $H_{PS}$. In the limit $W \gg \Omega$, the full single photon excited states $\ket{E_{i \pm}}$ are 
\begin{eqnarray}
\ket{E_{0 \pm}} &\equiv& \frac{1}{\sqrt{2}} \left[ \ket{1_l} \otimes \frac{\ket{0_r} + \ket{2_r}   }{\sqrt{2}} \otimes \ket{0_{Sl} 0_{Sr}}  \pm    \right. \\
& &  \left. \frac{\ket{0_l} + \ket{2_l}   }{\sqrt{2}} \otimes \frac{\ket{0_r} + \ket{2_r}   }{\sqrt{2}} \otimes \ket{1_{Sl} 0_{Sr}}          \right], \nonumber \\
\ket{E_{1 \pm}} &\equiv& \frac{1}{\sqrt{2}} \left[ \ket{1_l} \otimes \frac{-\ket{0_r} + \ket{2_r}   }{\sqrt{2}} \otimes \ket{0_{Sl} 0_{Sr}} \pm     \right. \nonumber \\
& &  \left. \frac{-\ket{0_l} + \ket{2_l}   }{\sqrt{2}} \otimes \frac{-\ket{0_r} + \ket{2_r}   }{\sqrt{2}} \otimes \ket{1_{Sl} 0_{Sr}}          \right]. \nonumber
\end{eqnarray}
Consequently, when a photon is lost from $\ket{L_0}$, the quantum system is placed in a superposition of $\ket{E_{0+}}$ and $\ket{E_{0-}}$, and will Rabi-flop at rate $\Omega$. However, photons in the shadow resonators rapidly decay, and the resulting $a_{SL}$ operation will return an $\ket{E_{0}}$ superposition to $\ket{L_0}$ and an $\ket{E_1}$ superposition to $\ket{L_1}$, without any additional phases accumulated in the process. Thus, photon loss errors are rapidly corrected in a manner which preserves superpositions of the two logical states; the energy conservation requirement enforced by $\delta \gg W \gg \Omega$ minimizes any excursions from the logical state manifold due to the error correction, and ensures that $\ket{E_0}$ only corrects to $\ket{L_0}$ and $\ket{E_1}$ only corrects to $\ket{L_1}$. However, a second photon loss in either qubit before correction occurs will lead to a logical error. Integrating out the shadow resonators, the ``repair" rate $\Gamma_R \of{\Delta E}$ for a process which changes the two-device system's energy by $\Delta E$ is given by:
\begin{eqnarray}
\Gamma_R \of{\Delta E}= \frac{4 \Omega^2 \Gamma_S }{ 4 \Omega^2 + 4 \of{ \Delta E + W + \frac{\delta}{2}}^2 + \Gamma_S^2}.
\end{eqnarray}
Here, $\Gamma_R$ is maximized when $\Delta E = -W - \delta/2$, which is precisely the energy of correcting a $\ket{E}$ state to its parent $\ket{L}$ state. Noting that there are an average of two photons in the circuit at any time and assuming $\delta \gg W$, we arrive at a net logical error rate from photon losses and off-resonant shadow resonator interactions of
\begin{eqnarray}\label{GEminus}
\Gamma_{E}^{X} &\simeq& 2  \Gamma_R \of{W + \frac{\delta}{2}} + \frac{2 \Gamma_P \of{2 \Gamma_P + \Gamma_R \of{ + W - \frac{\delta}{2} } }            }{  \Gamma_R \of{-W - \frac{\delta}{2}}              } \nonumber \\
\Gamma_{E}^{Y} &\simeq& \frac{2 \Gamma_P \of{2 \Gamma_P + \Gamma_R \of{ + W - \frac{\delta}{2} } }            }{  \Gamma_R \of{-W - \frac{\delta}{2}}              }.
\end{eqnarray}
Here, $\Gamma_{E}^{X}$ and $\Gamma_{E}^{Y}$ are the rates of random $\tilde{X}$ or $\tilde{Y}$ operations on a qubit in the circuit. In the limit $W \gg \Omega$ this is $4 \Gamma_P^2 / \Gamma_R \of{-W - \frac{\delta}{2}}$, which can be dramatically smaller than $\Gamma_P$. This rate describes the rate of random  $\tilde{X}$  (first term) and $\tilde{Y}$ (second term) operations, processes which can dephase a superposition of logical states or flip between them.

In FIG.~\ref{T1eff} we demonstrate the effectiveness of this protection against photon losses by numerically integrating the Lindblad equations \cite{gardinerzoller} for the system's density matrix $\rho$. Specifically, given a photon loss rate $\Gamma_P$, we have 
\begin{eqnarray}\label{lindblad}
\partial_t \rho &=& -\frac{i}{\hbar} \sqof{H,\rho} + \frac{\Gamma_P}{2} \sum_{j=L,R} \of{2 a_j \rho a_j^\dagger - \cuof{a_j^\dagger a_j,\rho}}  \nonumber \\
& & + \frac{\Gamma_S}{2} \sum_{j=L,R} \of{2 a_{Sj} \rho a_{Sj}^\dagger - \cuof{a_{Sj}^\dagger a_{Sj},\rho}}
\end{eqnarray}
As described in the figure caption, we can define two lifetimes for our logical states. The first, $T_{1L}$, is defined by initializing the system in either logical state and fitting the resulting decay to an incoherent mixture of the two logical states to an exponential decay law. The second, $T_{2L}$, is a dephasing time defined by initialization to the state $\of{\ket{L_0} \pm \ket{L_1}}/\sqrt{2}$ and fitting the expectation value of $\tilde{Z}_l \tilde{Z}_r$ to an exponential decay law. We note that $T_{2L}$ will  always be less than $T_{1L}$, as it is sensitive to both $\Gamma_{E}^X$ and $\Gamma_{E}^Y$ errors (\ref{GEminus}), where $T_{1L}$ is only sensitive to $\Gamma_{E}^{Y}$ processes. In both cases we neglect short time transient behavior (timescales less than $1/\Omega$), the effect of which is merged into an overall fidelity multiplier $F$. This stems from the fact that the system is measured, there is always a small chance of finding it outside of the logical state manifold, as the measurement may occur between a photon loss and its passive correction. This effect leads to a short-time dip in the expectation values of the logical operators $\tilde{X}$ and $\tilde{Z}_l \tilde{Z}_r$ after state initialization, where the error rate is $2 \Gamma_P$ for an interval of $\Delta t \approx \hbar /\Omega$ and slows down to the predicted rates in (\ref{GEminus}) after that (we neglect this short time behavior in our numerical fits to estimate lifetimes). However, if the measurement detects a $\ket{1}$ state, a subsequent measurement will capture the original (pre-loss) state with probability $P \simeq \Gamma_R / \of{ \Gamma_R + 2 \Gamma_P }$ due to the continuous passive error correction.

Finally, we should consider photon addition. An incoherent photon addition error can immediately lead to a logical error, since it takes $\ket{0} \to \ket{1}$ which is then rapidly converted to $\ket{2}$ by passive error correction, potentially enacting $\tilde{X}$. However, for modern, well-shielded experiments the available population of thermal photons is vanishingly small, and the random photon addition rate is two or more orders of magnitude less than the loss rate \cite{jinkamal2015}. This is thus unlikely to limit our logical state lifetimes.

\section{Error Suppression: Dephasing}

Having shown that our circuit is capable of rapidly correcting photon loss errors, we now demonstrate that the continuously applied many-body Hamiltonian $H_P$ (\ref{mainH}) required for error correction has the beneficial side effect of suppressing dephasing noise as well. Unlike the white noise of photon losses, dephasing noise has a power spectrum that is strongly frequency dependent, typically being comprised of $1/f$ and telegraph components \cite{martinisnam2003,ithiercollin2005,yoshiharaharrabi2006,bylandergustavsson2011,antonmuller2012,yangustavsson2013,paladinogalperin2014,omalleykelly2015}. The noise power spectra of these two sources are given by
\begin{eqnarray}\label{echo}
S_{1/f} &=& \frac{2 \pi S_0}{\omega}, \; S_{tel} = \frac{ \of{\Delta \omega_{10}}^2 \Gamma_{sw} }{\pi \of{\omega^2 + \Gamma_{sw}^2}   }.
\end{eqnarray}

If a system is continuously driven along $x$, the resulting interference between the effective Hamiltonian term $\eta \sigma^x$ and the fluctuating noise term $\delta z \of{t} \sigma^z$ can also strongly suppress phase noise \cite{martinisnam2003,ithiercollin2005,bylandergustavsson2011,yangustavsson2013,omalleykelly2015,yangustavsson2015}. When considering times $t > \eta^{-1}$, the average phase noise in this Rabi sequence is
\begin{eqnarray}\label{rabi}
\avg{\phi^2 \of{t} }^{\rm (Rabi)} \simeq \pi S \of{\eta} t,
\end{eqnarray} 
where $\eta$ is the Rabi frequency of the drive term ($2 \pi \times 35 {\rm MHz}$ in FIG.~\ref{T1eff}). This leads to exponential rather than Gaussian decay for both types of noise, and the noise suppression from a large $\eta$ can be dramatic. In our system, the large $W$ term will play exactly the same role, albeit with the noise strength $S_0$ increased by a factor of 4 relative to the single qubit noise measure, as we are working with two-photon states and there are two qubits experiencing noise. For a given $T_{2}^{\rm (echo)}$ from $1/f$ noise, the effective mixing time $T_{LZ}$ in our driven system can be vastly larger; for example, for single qubit $T_{2}^{\rm (echo)} = 10 {\rm \mu s}$ and $W = 2 \pi \times 35 {\rm MHz}$ we obtain $T_{LZ} \sim 2 {\rm ms}$. These results were confirmed to be qualitatively accurate by numerical noise simulations (see supplemental material). Since the $1/f$ $T_{2}^{\rm (echo)}$ scales as $1/\sqrt{S_0}$ and our Rabi-driven $T_{LZ}$ scales as $1/S_0$, a linear increase in $T_{2}^{\rm (echo)}$ from improved shielding or qubit design leads to a quadratic increase in $T_{LZ}$, just as in the photon loss channel. Similarly, for telegraph noise, $W \gg \Gamma_{sw}$ is readily achievable, and in this limit Rabi driving can outperform spin echo as well. To verify the prediction (\ref{rabi}), we simulated dephasing by averaging numerical simulations of randomly telegraph spectra. Within these simulations (included in the supplemental material), using published data from \cite{omalleykelly2015} we find a range of simulated $T_{LZ}$ values from $0.2 {\rm ms}$ for $\cuof{W = 2\pi \times 25 ,\Gamma_{sw}= 11.9 , \Delta \omega_{10} = 2 \pi \times 0.48  } {\rm MHz}$ up to $6 {\rm ms}$ for $\cuof{W = 2\pi \times 35 ,\Gamma_{sw}= 4.96 , \Delta \omega_{10} = 2 \pi \times 0.2  }$. We thus conclude that logical state lifetimes in the ms range are still achievable in the presence of realistic telegraph and $1/f$ noise sources.

We caution that our circuit offers no protection against true white noise dephasing (where $S \of{\omega}$ is constant at high frequency ranges), and increasing $W$ does not improve $T_{LZ}$ in this case. However, noise of this type is typically extremely weak and sometimes entirely absent in noise spectroscopies of modern superconducting qubit, with photon losses, flicker and telegraph noise dominating the error rate. Further, if white noise dephasing becomes a problem, it can be \textit{corrected} by constructing a three-device ring from our circuit, and implementing a passive variant of the three qubit phase flip code \cite{kapitchalker2015} alongside the passive photon loss correction.

Finally, we note that single qubit dephasing is not the only $z$ noise channel in our system, as two-body dephasing is also a concern. Specifically, flux noise through the coupling SQUID loop can lead to a fluctuating $\tilde{Z}_l \tilde{Z}_r$ term \footnote{Static $\tilde{Z}_l \tilde{Z}_r$ terms are not a serious concern, so long as they are known to the experimenter and small compared to $W$ and $\Omega$.}, though generally with a much smaller coefficient than the accompanying single qubit $\tilde{Z}$ terms. Because it commutes with $\tilde{X}_l \tilde{X}_r$ and mixes the two logical states this term is dangerous. Fortunately however, based on previous experiments with flux qubits (where $1/f$ flux noise through the qubit loop accounts for nearly all of the dephasing \cite{bylandergustavsson2011,yangustavsson2013,yangustavsson2015}) we expect this noise to be very weak at the symmetry point at which our device is operated, with a typical noise power $A \of{1 Hz } \simeq 1 \mu \Phi_0 / \sqrt{Hz}$. Assuming $1/f$ noise of this strength and the device parameters in the supplemental material ($E_J / E_C = 50$, with the two coupling SQUID junctions having energy $E_J = 2\pi \times 15 {\rm GHz}$), we obtain $T_{ZZ} \simeq 16 {\rm ms}$ as measured by an equivalent protocol to spin echo. More complex constructions can suppress this noise (such as through the introduction of a driven term $g \sqof{ \tilde{X}_l \of{1 + P_{r}^{1}} + \tilde{X}_r \of{1 + P_{l}^{1} } }$ added to $H_P$), if it ultimately becomes necessary.

\section{Logical Gates and Conclusion}

A simple universal two-qubit gate set can be implemented by combining single qubit rotations with the control-Z (CZ) operation. We let either $\tilde{X}$ operator play the role of logical $Z$ ($Z_L$). To enact single logical qubit rotations, we apply a finite length pulse involving combinations of a temporary phase shift for the signals which generate $W$ drive fields through the central SQUID (enacting $\tilde{Y}_l \tilde{Y}_r$, or $X_L$) and driving a single device resonantly at the $\ket{0} \leftrightarrow \ket{2}$ transition (enacting $\tilde{X}$, or $Z_L$). As the $g$ terms do not commute with $\tilde{Y}_l \tilde{Y}_r$, they may have to be briefly adjusted. An appropriately tuned sequence of these terms can rapidly enact arbitrary single-qubit rotations. Since the SQUID coupling can be driven fairly strongly (especially for short, highly tuned pulses), we do not expect these rotations to take significantly longer than in single qubit devices. To apply the CZ gate, we couple two of these qubit device pairs to each other, again through a driven SQUID coupling. Labeling the two device pairs (logical qubits) by $A$ and $B$, we simultaneously apply $\tilde{X}_{lA} \tilde{X}_{lB}$ through the coupling SQUID while applying $\tilde{X}_{lB}$ through the internal SQUID loop of the left qubit device of the $B$ pair. The sum of the two signals (which must be properly synchronized) run for an appropriate time enacts $I + \of{1 + Z_{LA}}\of{Z_{LB}}/2$, the logical CZ gate. Finally, our logical qubit could be measured along $\tilde{X}$ through a similar driven coupling to a resonator, analogously to the protocol proposed by Didier \textit{et al} \cite{didierbourassa2015}.

By considering a simple two-qubit circuit with driven couplings and two auxiliary lossy objects, we have demonstrated that passive error correction can lead to large improvements in qubit coherence against all common error channels with current technology. While our device is capable of only correcting or suppressing a single error at a time, it does so very rapidly, and permits simple and rapid logical gates between devices. We would like to develop a way to systematically integrate this logical bit into larger measurement based codes, and future study of hybrid QEC codes, where active and passive QEC methods work in concert, could be an extremely fruitful line of research.

\section{Acknowledgements}

We would like to thank Michel Devoret, Eric Holland, Jens Koch, Vadim Oganesyan, David Pappas and David Schuster for useful discussions. This work was supported by the Initiative for Theoretical Science at the Graduate Center of the City University of New York.

\bibliography{biblio,EC_bib,SLbib}

\newpage

\begin{center}
\textbf{Supplemental Information for A Very Small Logical Qubit}
\end{center}

\subsection{Device blueprint: $W$ terms and qubit-shadow resonator coupling}

In this section, we will explicitly derive the combination of drive signals needed to realize the Hamiltonians:
\begin{eqnarray}\label{HPS2}
H_P &=& - W \tilde{X}_l \tilde{X}_r  + \frac{\delta}{2} \of{P_{l}^{1} + P_{r}^{1}}. \\
H_{PS} + H_{S} &=&  \of{W + \frac{\delta}{2} }  \of{ a_{Sl}^\dagger a_{Sl} +  a_{Sr}^\dagger a_{Sr} } \\ & &+ \Omega \of{a_l^\dagger a_{Sl}^\dagger + a_r^\dagger a_{Sr}^\dagger + {\rm H.c.} }. \nonumber
\end{eqnarray}
Here, as in the main text, we let $P_{k}^{n} \equiv \ket{n_k} \bra{n_k}$ be the projector onto all states with $n$ photons in object $k$ (and any number of photons in the other parts of the circuit). We further define $\tilde{X}_l \equiv \of{ a_l^\dagger a_l^\dagger + a_l a_l}/2$ and $\tilde{Z}_l \equiv P_l^2 - P_l^0$ (and similarly for $r$), where $a_{l}$ annihilates a photon in the left device. The derivation of (\ref{HPS}) will be for transmon \cite{kochyu,houckkoch} qubits, but the generalization to flux or fluxonium qubits is straightforward. We consider a pair of superconducting transmon qubits which share a common (bridged) ground, as shown in FIG.~\ref{circuitfig}. We label the two transmons by $l$ and $r$, and design or flux tune them so that their excitation energies $\omega_{l}$ and $\omega_r$ differ substantially (a GHz or more). The qubits are coupled to each other by a SQUID, with flux biases $\Phi_{1} \of{t}$ and $\Phi_{2} \of{t}$ threaded through the larger (inner) and smaller (SQUID) loops, respectively. We let the two junction energies be equal to $E_{Ji}$; the coupling Hamiltonian is then equal to:
\begin{eqnarray}
H_i = - E_{Ji} \sqof{ \cos \of{\delta \phi + \Phi_1 \of{t} } + \cos \of{\delta \phi + \Phi_1 \of{t} + \Phi_2 \of{t} }  } \nonumber
\end{eqnarray} 
We now let:
\begin{eqnarray}
\Phi_{1} \of{t} = -\frac{\pi}{2} + f \of{t}, \; \Phi_{1} \of{t} = \pi - 2 f \of{t},
\end{eqnarray}
where $f \of{t}$ is a rapidly oscillating flux signal. Assuming the two junctions are equal in energy, this reduces to
\begin{eqnarray}
H_i = - 2 E_{Ji} \cos \delta \phi \sin f \of{t}.
\end{eqnarray}
If the two Josephson energies are unequal, we will have an additional term which scales as $\of{E_{J1} - E_{J2} } \sin \delta \phi \cos f \of{t}$; all terms which come out of this coupling are rapidly oscillating and can be dropped. We choose to include two junctions here rather than the single junction used in \cite{kapit2015} so that we can make $E_{Ji}$ large without worrying about instabilities from the off-resonant $\sin \delta \phi$ terms. We now wish to choose $f \of{t}$ so that the third order term from expanding the sine produces the target signals:
\begin{eqnarray}\label{freqtarget}
f \of{t}^{3} &=& 3 \alpha^{3} \sqof{ \cos{2 \of{\omega_l - \omega_r} t} + \cos \of{2 \of{\omega_l + \omega_r - \delta} t} } \nonumber \\
& & + { \rm (r.o.)}; \quad f \of{t}^{1} = { \rm (r.o.).}
\end{eqnarray}
Here, ``(r.o.)" is short for rapidly oscillating; e.g. any portions of the signal which are far from any resonant transitions that can be induced by the operator $\cos \delta \phi$. To provide a realistic example, if $\cuof{\omega_l,\omega_r,\delta}= 2\pi \times \cuof{4.5,6.5,-0.35}$ GHz, we can choose
\begin{eqnarray}
f \of{t} &=& \alpha \left[ \cos \of{\frac{2 \omega_l + 6 \omega_r - 4 \delta}{5} t} \right. \\
 & & \left. + \cos \of{\frac{6 \omega_l - 2 \omega_r - 2 \delta}{5} t}  \right]. \nonumber
\end{eqnarray}
This combination is detuned from all unwanted combinations by at least a GHz, and requires drive frequencies of only 7.72 and 5.86 GHz, well within normal experimental operation ranges. One could of course realize these the terms in (\ref{freqtarget}) at order $\alpha$ using a simple direct drive at those target frequencies, but as the four-photon term occurs requires frequencies in the 15-25 GHz range it could be difficult to achieve with commonly available microwave hardware.

To reduce this complex driven interaction to a simple, rotating frame Hamiltonian, we begin by observing that the phase operators $\cos \phi$ and $\sin \phi$ for the first three levels can be written as:
\begin{eqnarray}\label{opdefs}
\sin \phi =  \begin{pmatrix}
0 & S_{01} & 0 \\
S_{01}^* & 0 & S_{12} \\
0 & S_{12}^* & 0
\end{pmatrix}, \; \cos \phi = \begin{pmatrix}
C_{00} & 0 & C_{02} \\
0 & C_{11} & 0 \\
C_{02}^* & 0 & C_{22}
\end{pmatrix}.
\end{eqnarray}
Here, the $S$ and $C$ coefficients can be determined numerically by diagonalizing the single qubit Hamiltonian. If we transform to the rotating frame via the transformation 
\begin{eqnarray}
\ket{\Psi} \to e^{ i \sqof{ \of{\omega_l - \delta/2 } \of{a_{l}^\dagger a_l}  +  \of{\omega_r - \delta/2 } \of{a_{r}^\dagger a_r} } t } \ket{\Psi},
\end{eqnarray}
then neglecting all rapidly oscillating terms the state $\ket{\Psi}$ evolves under the effective primary Hamiltonian $H_P$: 
\begin{eqnarray}
H_P &=& - W \tilde{X}_l \tilde{X}_r  + \frac{\delta}{2} \of{P_{l}^{1} + P_{r}^{1}}. 
\end{eqnarray}
Here, $W = - E_{Ji} \alpha^3 \abs{C_{02}}^2 / 4 $, which can be in the $10-40$ MHz range for realistic parameters.

We now turn to the resonators, which we label $Sl$ and $Sr$ (the $S$ label denotes a shadow object, as discussed in \cite{kapithafezi2014,kapitchalker2015}), with energies $\omega_{Sl}$ and $\omega_{Sr}$. These resonators could also be (intentionally lossy) qubits; the analysis in here proceeds identically in either case. As the target frequencies for the two-photon drive are much lower than for the four-photon term in $W$, we can use a simple direct drive. If the flux biases through the qubit-resonator loops are $\pi/2 + g_{Sk} \of{t}$ and $\pi - 2 g_{Sk} \of{t} $ as outlined above (where $k=l,r$), we can let
\begin{eqnarray}
g_{Sk} \of{t} = \beta \cos \sqof{\of{\omega_{k} + \omega_{Sk} - W - \delta}}t
\end{eqnarray}
Plugging this into the interaction terms and enacting a similar rotating frame transformation on the resonators to cancel the time dependence, our resulting qubit-resonator Hamiltonian is:

\begin{eqnarray}
H_{PS} + H_{S} &=&  \of{W + \frac{\delta}{2} }  \of{ a_{Sl}^\dagger a_{Sl} +  a_{Sr}^\dagger a_{Sr} } \\ & &+ \Omega \of{a_l^\dagger a_{Sl}^\dagger + a_r^\dagger a_{Sr}^\dagger + {\rm H.c.} }. \nonumber
\end{eqnarray}
Our final device Hamiltonian is simply $H = H_P + H_{PS} + H_{S}$ (\ref{HPS2}), as desired.

\subsection{Quasiparticle tunneling}

One potentially concerning source of error is quasiparticle tunneling across a coupling junction. In ordinary transmons, random tunneling from non-equilibrium quasiparticle populations is one of the dominant error channels after $T_1$ exceeds $100 {\rm \mu s}$ \cite{catelanikoch2011,wanggao2014,jinkamal2015}, leading to photon losses through the $\sin \phi/2 $ operator. We expect these errors to occur in our circuit as well, along with an additional error channel from a quasiparticle tunneling from one qubit to the other across the coupling SQUID. The corresponding operator $\sin \of{\frac{\phi_l - \phi_r}{2}}$ for this process has components which act as $a_l$, $a_l^- \tilde{Z}_r$ and $a_l a_r a_r$, corresponding to a single photon loss, a combination of loss and dephasing and a three photon loss, respectively (and equivalent operators with $l$ and $r$ switched). The first process has the largest matrix element, but is just an ordinary single photon loss (contributing to a small increase in $\Gamma_P$) and thus will be rapidly corrected. The second and third processes are more serious, as they are multi-qubit errors that can induce logical state transitions when corrected by engineered dissipation. Fortunately however, the matrix elements for these terms are very small. Assuming $E_J / E_C = 50$, the squared matrix elements for the $a_l \tilde{Z}_r$ and $a_l a_r a_r$ terms are reduced by factors of 0.004 and 0.002, respectively, compared to the bare matrix element for a single photon loss. These terms are thus unlikely to place significant limits on $T_{1L}$. However, if single qubit $T_1$'s were long enough for these terms to become a real problem, they can always be eliminated by adding a second, identical coupling SQUID, with a flux bias of $2\pi$ relative to the original SQUID. This creates perfect destructive interference for quasiparticle tunneling without changing the basic coupling structure \cite{catelanikoch2011}, eliminating this error channel. 

\subsection{Phase noise}

In this section, we describe the effect of estimating the effect of phase noise in the circuit. To model the effect of phase noise (random $\tilde{Z}$ operations), it is sufficient to consider a single spin 1/2 degree of freedom with Hamiltonian $H = - W \sigma^x$ experiencing noise through a fluctuating term $\delta z \of{t} \sigma^z$, since $\tilde{Z}$ errors do not change photon number and thus cannot mix with the $\ket{1}$ state for either transmon. By initializing the spin in $\sigma^x = 1$  and averaging over noise patterns $\delta z \of{t}$ randomly generated to obey a given noise power spectrum, we can obtain a lifetime $T_2$ for the $\sigma^x$ eigenstates; the resulting logical state lifetime $T_L$ in our 2-qubit device will be half of this since there are two error channels. Phase noise with a white spectrum ($S \of{\omega}$ is constant) produces simple exponential decay, and needs no simulation.

\begin{figure}
\includegraphics[width=3.25in]{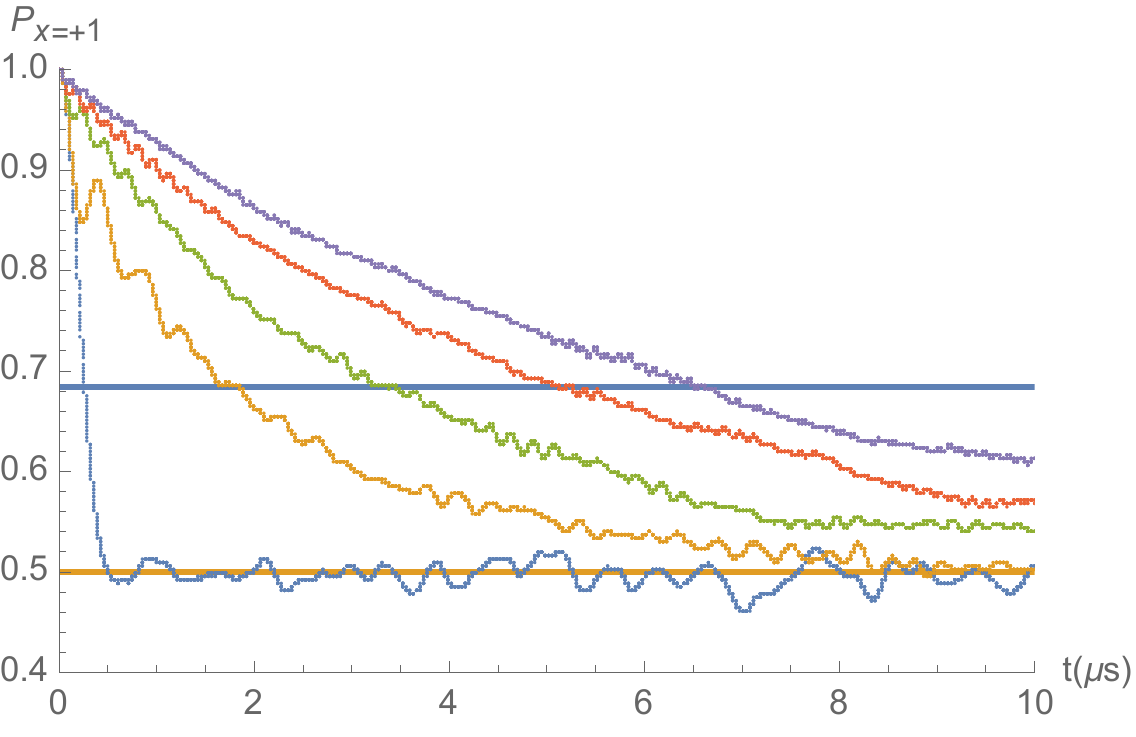}
\includegraphics[width=3.25in]{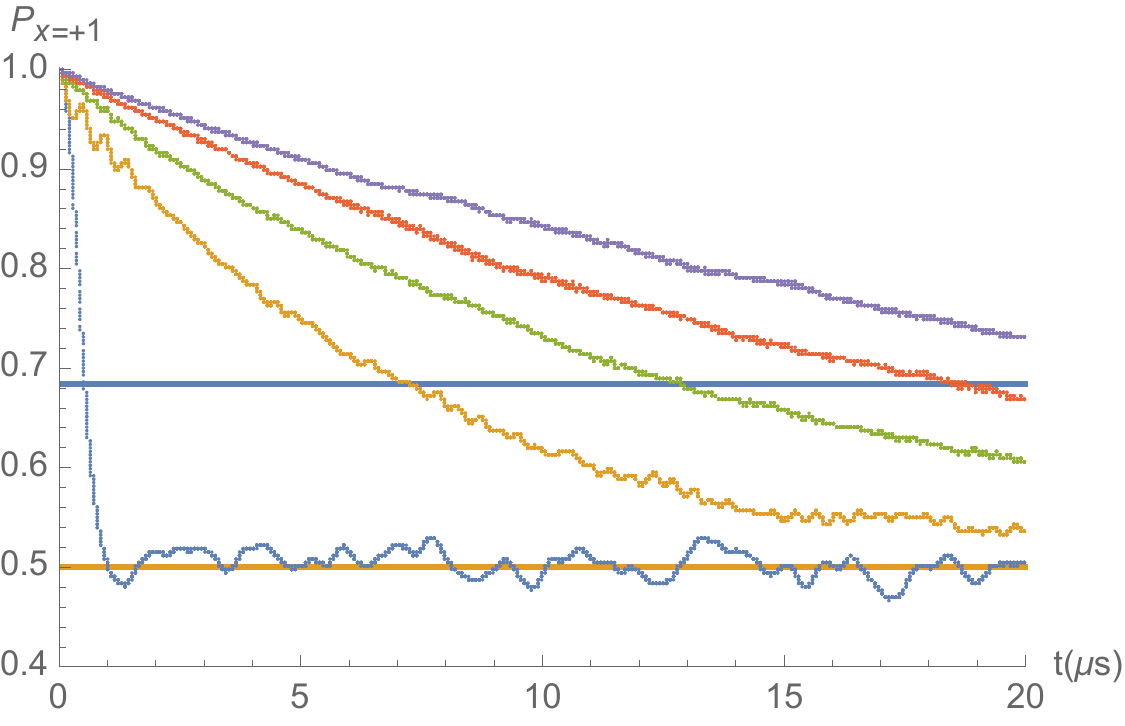}
\caption{Simulations of $1/f$ noise along $z$, suppressed by a drive term $H = W \sigma^x$, with (in order of increasing lifetime) $W = 2\pi \times \cuof{0,1,2,3,4}$ MHz. The flat blue line demarcates $1/e$ for ease of extracting the lifetime. In the top plot, the noise strength chosen so that $T_{2}^{(echo)} = 1 \mu s$, and in the bottom plot we rescale the noise strength by $1/4$ to obtain $T_{2}^{(echo)}= 2 \mu s$. The continuous drive fields lead to a simple exponential decay law for the initial qubit state (rather than Gaussian for undriven Ramsey or echo decays), and as discussed in the text, a linear increase in the drive term $W$ leads to a linear increase in $T_2$, while a linear decrease in the noise power leads to a quadratic increase in $T_2$. Each curve is the average of 900 randomly generated noise patterns.}\label{1overf_fig}
\end{figure}

To verify the prediction $T_{2}^{\rm Rabi} \simeq \of{ \pi S \of{W}}^{-1}$ of $1/f$ noise suppression by Rabi driving, we simulated randomly generated noise traces with an average $1/f$ spectrum, and averaged the expectation value $\avg{P_{\sigma^{x}=1}}$ over these traces with a variable driving Hamiltonian $H = W \sigma^x$. As shown in FIG.~\ref{1overf_fig}, the scaling form $T_{2}^{\rm Rabi} \simeq \of{ \pi S \of{W}}^{-1}$ is in good quantitative agreement with the numerically simulated evolution, underestimating the lifetime by less than 20\% and accurately capturing the scaling with $W$ and $S_0$.

For telegraph noise the situation is somewhat more complex, as the noise spectrum is defined by two parameters, the energy difference $\Delta \omega_{10}$ created by the noise term, and $\Gamma_{sw}$, the incoherent switching rate between the two states. We considered the noise spectrum of a single telegraph fluctuator, which shifts the single photon energy of a qubit by $+\Delta \omega_{10}$ when in the ``on" state, does nothing in the ``off" state, and will randomly switch between the two states with switching rate $\Gamma_{sw}$. To study the effect of such a flucuator, we computed lifetimes by sampling over 250 points in the box defined by $W \in 2\pi \times \cuof{ 10,  37.5} {\rm MHz}$, $\Delta \omega_{10} \in 2\pi \times \cuof{ 0.1, 0.55} {\rm MHz}$ and $\Gamma_{sw} \in \cuof{4,22} {\rm MHz}$. For each combination of $\cuof{W, \Delta \omega_{10}, \Gamma_{sw}}$ we simulated the evolution under 900 randomly generated noise traces, and fit the resulting curve to an exponential decay law $\avg{\sigma^x \of{t}} = e^{-t / T_{2}}$. We then numerically fit the data to the form $a W^{b} \Delta \omega_{10}^{c} \Gamma_{sw}^{d}$ to arrive at the expression:
\begin{eqnarray}
T_{2} \of{W, \Delta \omega_{10}, \Gamma_{sw}} \simeq  \frac{ 2.30 W^{1.98}}{\Delta \omega_{10}^{2} \Gamma_{sw}^{1.07}} \simeq  \frac{ 2.30 W^{2}}{\Delta \omega_{10}^{2} \Gamma_{sw}}
\end{eqnarray}
As described in the main text, for realistic device parameters taken from contemporary experiments, the resulting $T_{2}$ for our two-qubit device (which is one eighth the above value as there are two noise sources and the value $\Delta \omega_{10}$ should be doubled for two-photon states) could easily be in the range of a few ms. The protection against telegraph phase noise from continuous driven evolution is thus comparable to the protection against photon losses via resonant energy transfer to the shadow resonators.

\subsection{Two body phase error suppression}

As described in the text, two-body phase errors, where flux noise through the Josephson coupling randomly enacts $\tilde{Z}_l \tilde{Z}_r$, may become a limiting factor for long-lived qubits. Should it become necessary, we here propose a method of suppressing them which will not interfere with passive error correction. We wish to add to $H_P$ a term of the form:
\begin{eqnarray}
g \sqof{ \tilde{X}_l \of{1 + P_{r}^{1}} + \tilde{X}_r \of{1 + P_{l}^{1} } }
\end{eqnarray}
The reason for choosing this more complex term rather than $\tilde{X}_l + \tilde{X}_r$ is to ensure that the rotating frame energy cost of a single photon loss is $W + \delta/2$ for either $\tilde{X}$ eigenstate. If the $P_{l/r}^1$ terms are absent and there is an energy mismatch, a superposition of the two logical states will rapidly dephase whenever an error occurs from a single photon loss, but if they are present no additional phases will accumulate while the photon loss error is passively corrected. Though complex, such a term could thus be valuable, as it anticommutes with $\tilde{Z}_l \tilde{Z}_r$ and thus suppresses low frequency errors in that channel.

We can generate these $g$ terms perturbatively using additional drive fields. We consider a modification of the circuit so that the qubits are split transmons, and have an internal SQUID loop that can be driven to enact single qubit operations. We then add a signal component $\kappa_1 \cos \frac{\of{\omega_l - \delta/2}}{2} t$ to the coupling SQUID, adding a term $- E_{Ji} \sin{\Delta \phi} \cos \sqof{\kappa_1 \cos \frac{\of{\omega_l - \delta/2}}{2} t}$ to $H_P$. The nontrivial rotating frame contribution from this term is $ E_{Ji} \kappa_{1}^{2} \sin \phi_l \cos \phi_r \cos \of{\of{\omega_l - \delta/2}t} / 4$. Finally, we apply three frequencies through the left transmon's internal SQUID loop, coupling to $\cos \phi_l$ and taking the form:
\begin{eqnarray}
f_l \of{t} = \sum_{k = -1}^{1} \kappa_{2,k}  \cos \sqof{\of{\omega_l - \frac{\delta}{2} - k \nu} t}
\end{eqnarray}
Here, $\nu$ is chosen to far detune these signals from any other relevant transitions (alternative sets of signals can be chosen here, so long as one remains at $\omega_l - \delta/2$ and the other two frequency shifts sum to zero). This adds two terms to the system Hamiltonian (along with other terms we neglect for ultimately being rapidly oscillating):
\begin{eqnarray}
H_\kappa &=& -\of{\kappa_{2,0} E_{Jl} +  \of{\kappa_1^2 /4} E_{Ji} \cos \phi_r } \times \\
& & \sin \phi_l \cos \sqof{ \of{\omega_l - \frac{\delta}{2} }t} \nonumber \\
  & & + E_{Jl} \frac{\kappa_{2,0}^2+ 2 \kappa_{2,-1} \kappa_{2,1}}{4} \cos \phi_l \cos \sqof{ \of{2 \omega_l - \delta} t }. \nonumber
\end{eqnarray}
The first line of terms describes an off-resonant single photon drive field, which we can eliminate perturbatively. Specifically, and taking the fourth level into account as a perturbative correction, the above Hamiltonian becomes
\begin{eqnarray}
H_\kappa &=& \frac{\of{ \kappa_{2,0} E_{JI} + \of{\kappa_1^2 /4} E_{Ji} \cos \phi_r}^{2}}{2 \delta} \of{ c_1 \tilde{X}_l + c_2 \tilde{Z}_l} \nonumber \\
& & + c_3 \tilde{X}_l. \nonumber
\end{eqnarray}
For large $E_J / E_C$ the $c_2$ term turns out to be small enough to ignore. We now rewrite the undriven $\cos \phi_r$ as a matrix operator in our three-level basis (keeping only the diagonal elements), and by carefully choosing the four $\kappa$ coefficients, we can always arrive at
\begin{eqnarray}
H_\kappa \simeq g \of{1 + P_r^1} \tilde{X}_l,
\end{eqnarray}
where the approximate equality indicates that we have dropped terms small static terms with coefficients an order of magnitude weaker than the desired terms. If we now add an equivalent set of signals for the right transmon, we will recover the $g$ terms in the device Hamiltonian, protecting the circuit against two-body phase errors in the same way that the $W$ terms protect against single-qubit phase errors.
 
\subsection{``Leakage-free" error correction}

As seen in FIG.~2 of the main text, as the photon loss rate $\Gamma_P$ decreases errors induced by the passive error correction itself begin to limit lifetime. The two processes which cause this limit are (a) excursions from the logical state manifold due to off-resonant transitions from a one-photon error state to a state where $\tilde{X}_l \tilde{X}_r = -1$, and (b) off-resonant photon addition from the shadow qubit coupling, which sends either logical state to a one-photon state and then promptly back to the logical state manifold through error correction, leading to a random $\tilde{X}$ operation. Process (a) limits both $T_{1L}$ and $T_{2L}$, whereas process (b) only limits $T_{2L}$. Either way, we imagine that these error processes can be suppressed by replacing the monochromatic tones which drive the qubit-shadow object coupling with more complex periodic sequences, specially tuned to eliminate the off-resonant transitions. Previous experiments in superconducting qubits have shown that such pulse shaping can suppress leakage (unwanted mixing with $\ket{2}$ states) by many orders of magnitude in quantum gates \cite{motzoigambetta2009,chenkelly2015}, and were we to use the same methods to eliminate mixing with $\ket{1}$ states and the $\tilde{X}_l \tilde{X}_r = -1$ manifold, we would arrive at lifetimes which scale as $4 \Gamma_P^{2} / \Gamma_R$ even at very small $\Gamma_P$, a large improvement over the result in FIG.~2 of the main text. We expect that such methods will be necessary to achieve $T_{L}$ values in the 10 ms range using this architecture, but the precise details of their implementation are beyond the scope of this paper and unimportant for reaching the low ms range using current technology. 
 
\end{document}